\documentclass[a4paper,fleqn]{cas-dc}

\usepackage[authoryear,longnamesfirst]{natbib}
\usepackage{comment}
\usepackage[vlined,linesnumbered,ruled,boxed]{algorithm2e}
\usepackage{tikz}
\newcommand*\circled[1]{\tikz[baseline=(char.base)]{
            \node[shape=circle,draw,inner sep=2pt] (char) {#1};}}

\def\tsc#1{\csdef{#1}{\textsc{\lowercase{#1}}\xspace}}
\tsc{WGM}
\tsc{QE}
\tsc{EP}
\tsc{PMS}
\tsc{BEC}
\tsc{DE}

\newcommand{\eg}{{\it e.g., }}
\newcommand{\etal}{{\it et~al. }}
\newcommand{\ie}{{\it i.e., }}

\begin{document}
\let\WriteBookmarks\relax
\def\floatpagepagefraction{1}
\def\textpagefraction{.001}

\shorttitle{Resource allocation across serverless fog federation}

\shortauthors{Hussain et~al.}

\title [mode = title]{Resource Allocation of Industry 4.0 Micro-Service Applications across Serverless Fog Federation}                      
\tnotemark[1,2]



%
\author[1]{Razin Farhan Hussain}



\ead{razin@trycycledata.com}



\affiliation[1]{organization={TryCycle Data Systems},
    addressline={1296 Carling Ave}, 
    city={Ottawa},
    postcode={K1Z 7K8}, 
    country={Canada}}

\author[2]{Mohsen Amini Salehi}

\ead{mohsen.aminisalehi@unt.edu}
\ead[URL]{https://hpcclab.org/}


\affiliation[2]{organization={University of North Texas (UNT)},
    city={Denton},
    postcode={76203}, 
    state={Texas},
    country={U.S.A.}}



\begin{abstract}
The Industry 4.0 revolution has been made possible via AI-based applications (\eg for automation and maintenance) deployed on the serverless edge (aka fog) computing platforms at the industrial sites---where the data is generated. Nevertheless, fulfilling the fault-intolerant and real-time constraints of Industry 4.0 applications on resource-limited fog systems in remote industrial sites (\eg offshore oil fields) that are uncertain, disaster-prone, and have no cloud access is challenging. It is this challenge that our research aims at addressing. We consider the inelastic nature of the fog systems, software architecture of the industrial applications (micro-service-based versus monolithic), and scarcity of human experts in remote sites. To enable cloud-like elasticity, our approach is to dynamically and seamlessly (\ie without human intervention) federate nearby fog systems. Then, we develop serverless resource allocation solutions that are cognizant of the applications' software architecture, their latency requirements, and distributed nature of the underlying infrastructure. We propose methods to seamlessly and optimally partition micro-service-based application across the federated fog. Our experimental evaluation express that not only the elasticity is overcome in a serverless manner, but also our developed application partitioning method can serve around 20\% more tasks on-time than the existing methods in the literature.

\end{abstract}




\maketitle

\section{Introduction}\label{intro}
The Industrial Revolution brought about rapid changes in operations by incorporating state-of-the-art technologies. However, various solutions must be synchronized and adapted accordingly. This rapid shift is prevalent, especially in remote sites. Nonetheless, processing emerging operational data and smart applications on available computing platforms can be challenging in harsh operational environments, which motivates our research work. Therefore, we explore the stochastic behaviours and in-depth structure of Industry 4.0 applications and address the challenges of modern computing platforms in the following sections.
\subsection{Overview and Motivation}
Industrial systems are rapidly shifting from human-controlled processes towards closed-loop serverless control systems that process various types of applications to manage industrial operations autonomously. Particularly at remote sites, such as offshore oil and gas fields (\cite{hussain2022iot}), space stations (\cite{aume2022trackink}), submarines and underwater robots (ROVs) (\cite{kabanov2022marine}), the Industry 4.0 paradigm shift demands systems to serverlessly process emerging data-driven and latency-sensitive applications under harsh operational environment where there is limited/no access to the cloud services, and human resources are scarce and not computer literate. Realizing these systems mandates addressing challenging research questions to enable robust, latency-aware, and serverless processing of the applications on alternative computing platforms (\cite{wang2019big,8291112, cai2019iot}) operating atop low-latency wireless communication systems (\cite{gao2023sparklink}).

To overcome the lack of reliable access to cloud servers, making use of the fog computing systems (\cite{mattia2023p2pfaas}) in remote industrial sites has become a common practice. Nonetheless, these fog systems inherently suffer from the lack of elasticity (\cite{nguyen2020elasticfog}) and fail to handle workload spikes often occur due to unpredictable disasters that the remote industrial sites are prone to (\cite{chiou2022robot}). This lack of elasticity and resource scarcity curbs the excessive use of compute-intensive (\eg AI-based) solutions at the fog platform level. In practice, managing emergency situations demands lightweight and explainable solutions that operate fast and do not impose extra burden to the fog system. An exemplar use case is a remote (offshore) oil field where, upon detecting an oil spill, the following coordinated activities must be processed within a short period of time: (A) Drones must be dispatched for more granular investigation; (B) Emergency teams must be notified; and (C) High-end simulations must be conducted for purposes like predicting the oil spill expansion direction, and staff evacuation. 

To establish an Industry 4.0 system in a remote site that is robust against such unpredictabilities, we explore the challenging idea of augmenting the processing capability of the local fog via dynamically federating it with nearby fog systems (\eg mobile datacenters (\cite{baburao2023novel})), thereby, providing cloud-like serverless elasticity for Industry 4.0 applications. This challenge stems from the fact that modern industrial applications are often cloud-native, and are not originally designed to reap the benefits of wirelessly-connected autonomous fogs. \emph{There is an infrastructural gap to adapt these applications to the federated fog environment, and this gap is what this research aims at filling}. More specifically, the challenge is how to establish the notion of serverless such that the applications can seamlessly take advantage of the dynamically formed federated fog system? To overcome this challenge, we need to deal with two aspects of the federated environments: 

\noindent\textit{(a) Characteristics of the federated fog environment}:
The federated fog environment is prone to the uncertainties stem from the unreliable communication between fog system, and heterogeneous computing across them. Therefore, these uncertainties can potentially affect the latency constraint of Industry 4.0 applications across the federation. Failure to dealing with these uncertainties can potentially hurt the robustness of remote site instead of helping it (\cite{salehi2016stochastic}).

\begin{figure}[h!]
	\centering	
	\includegraphics[width=0.47\textwidth]{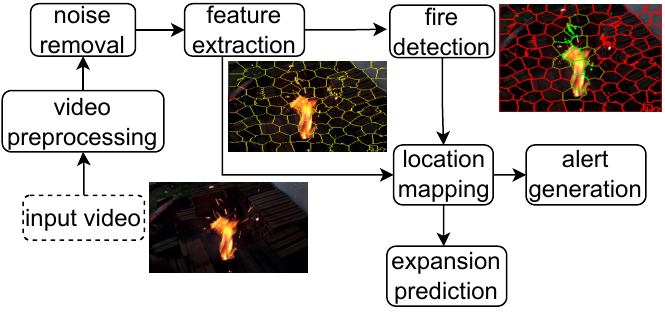}
	\caption{Workflow of micro-services needed for the ``fire detection'' application. The application needs to seamlessly make use of federated fog to complete on time and prevent any potential damage.\label{fig:workflowMicro}}
\end{figure}

\noindent\textit{(b) Characteristics of Industry 4.0 applications}: Most of the current Industry 4.0 applications function based on  Machine Learning (ML-based) and typically follow the micro-service-based software architecture (\cite{jwo2022data,aberle2022microservice}) where a workflow of micro-services (\cite{dragoni2017microservices}) have to be completed within a deadline. There are also legacy applications with monolithic architecture that are inflexible and have to be allocated in an atomic manner (\ie cannot be partitioned into modules) (\cite{calderon2018integration}). Unlike monolithic applications, micro-services are composed of several independent services that can be deployed separately. For instance, as depicted in figure \ref{fig:workflowMicro}, a ``fire safety" application is essentially a workflow of micro-services that includes services for capturing surveillance video content, pre-processing the captured content, noise removal, feature extraction, fire detection, location mapping, alert generation, and expansion prediction. Hence micro-services can be constrained by certain factors (e.g., location and data). For example, a micro-service that reads data from surveillance camera has to be located on the server connected to the camera. Such constraints have to be considered in allocation and partitioning of the Industry 4.0 applications.


It is crucial to have a lightweight and efficient solution that can handle sudden surges in demand without adding computing stress to the overall system, especially in unpredictable emergencies with limited computing resources. In sum, an ideal platform for remote sites should be: (i) serverless in the sense that it can seamlessly supply resources for both monolithic and micro-service applications across the federation; and (ii) simple, robust, and effective against the network and computing uncertainties such that it can fulfill the  latency (deadline) constraints of the applications. 

\subsection{Problem Statement and Contributions}
\begin{figure}[h!]
	\centering	
	\includegraphics[width=0.55\textwidth]{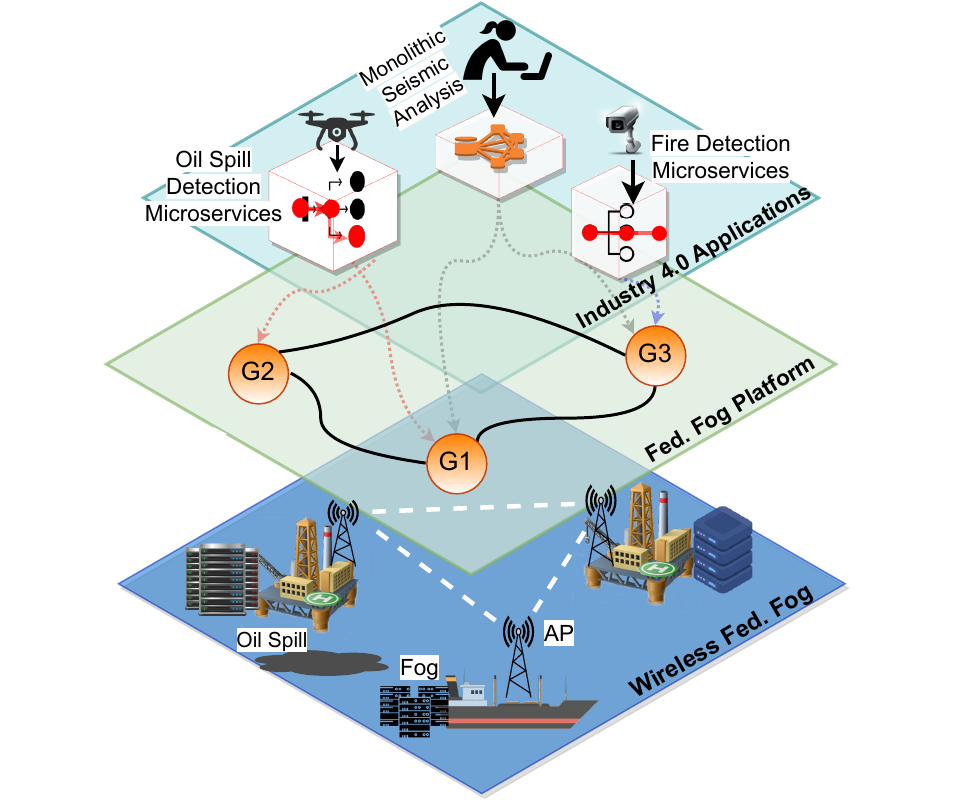}
	\caption{A multi-layer view of the serverless fog federation infrastructure in the context of offshore smart oil and gas industry. Micro-service and monolithic Industry 4.0 applications seamlessly run across the federated fog systems by means of a platform, deployed on each gateway ($G_i$) representing a fog system. \label{fig:layerFog}}
\end{figure}

As depicted in Figure \ref{fig:layerFog}, for the use case of remote smart oil fields, a federated fog computing can be formed via wireless communication between nearby fog servers. Considering each fog system $i$ as a node (gateway) $G_i$ and the wireless links between them as edges, we can conceptually model the federated fog in form of a graph that the federated fog platform operates upon. It is noteworthy that each fog system is configured with a different set of resources, \ie there is a heterogeneity across fog systems. In this system, uncertainty exists in both the communication and computation times of the same task running across the federation. Accordingly, to meet the latency constraints (deadline) of the applications, the federated fog platform should be cognizant of these communication and computing heterogeneity and uncertainties, in addition to the applications' characteristics and latency requirements. 

Considering the heterogeneity across fog systems, in our prior work~(\cite{hussain2020analyzing}), we explored and identified the stochastic execution behavior of several Industry 4.0 applications across various type of machines. In another study~(\cite{hussain2019federated}), we investigated resource allocation strategies for the monolithic applications across the fog federations. 
Nonetheless, many real-world operational processes are carried out by means of applications with a workflow (Directed Acyclic Graph---DAG) of micro-services, where each micro-service is an independent entity that communicates its output data as the input to the next micro-service in the graph. Development and deployment of such applications on the federated fog entails the user's involvement in the topological and uncertainty details of the underlying infrastructure to ensure that the whole application is complete within its deadline. To establish the notion of serverless, the federated fog platform has to transparently take care of these details and allocate the application such that application's deadline is met.  

Micro-service DAGs offer the opportunity to perform the allocation at the micro-service level. That is, the platform can partition the workflow graph into subgraphs and allocate them across the federation. In this case, for a given workflow topology, and specific federation uncertainties, the two following questions have to be addressed: (1) How to partition the micro-service workflows, so that its deadline can be met? (2) How to allocate computing resources to the partitioned micro-services across fog federation, so that it has the highest likelihood of meeting its deadline?

To address these questions, we propose a load-balancing approach within the federated fog platform that is aware of both the application's software architecture, its deadline, and the underlying execution platform's characteristics. As a result, it enables cloud-like serverless behavior via dynamically and seamlessly allocating the applications across the federation. 
In summary, this paper makes the following contributions:
\begin{itemize}
    \item Developing a probabilistic method (called \textit{ProPar}) to optimally partition micro-service workflows across the fog federation in a transparent and serverless manner.
    
    \item Developing a resource allocation method based on the Bayesian statistics that is cognizant of the application deadline and efficiently assigns partitioned workflows across the federation.
    
    \item Evaluating and analyzing the partitioning and the resource allocation methods across the fog federation for Industry 4.0 applications.
\end{itemize}

The outcomes of this research lays the foundation upon which solution architects and industry experts can focus only on the business logic of their applications--either micro-service-based or monolithic--and leave the allocation details of their applications and handling the demand surge to the serverless fog federation platform. The rest of the paper is organized in the following manner. Section~\ref{relatedwork} presents related works as background studies. Section~\ref{sysModel} represents the system model. Section~\ref{partitioning} states the partitioning method of micro-service workflow across fog federation, whereas section~\ref{resourceAllocation} represents the resource allocation method for Industry 4.0 application. Section~\ref{result} demonstrates experimental setup, baseline techniques, and experimental results. Finally, section~\ref{concl} concludes the paper with some future avenues for exploration.
\section{Background and Prior Studies}\label{relatedwork}
Serverless computing is a computing model that enables developers to deploy and run applications without managing the underlying infrastructure \cite{cinque2023real}. This new paradigm has become popular due to the rise and adoption of containerization and micro-services in Industry 4.0. Applications are designed to be divided into small, self-contained functions that operate independently and can respond to specific events. These functions are short-lived, stateless, and highly scalable, making them ideal for handling unpredictable workloads and traffic spikes \cite{patros2021toward}. Overall, serverless computing offers reduced costs, increased agility, and faster time-to-market. Various research works (\cite{oluyisola2022designing,ammar2022implementing,laskar2022smart,al2022well,dai2022uav}) have been undertaken for the development and integration of smart applications into the industrial sectors utilizing serverless technology. 
\subsection{Resource Allocation in Industry 4.0}
The resource allocation in Industry 4.0 is challenging due to dynamic user demands and limited resources. Effective resource allocation and management must adapt to the changing needs. Industrial IoT generates enormous data that needs fast processing that is enabled by fog computing---bridging the gap between cloud and IoT devices. In a recent study \cite{atiq2023reliable} Haseeb \etal propose a cost-effective resource allocation and management strategy considering latency and energy efficiency. Authors of this work introduce a framework, called R2AM, to manage resources in transportation IoT using fog computing. Data from IoT devices is queued for storage and processing, while fog nodes are sorted based on their processing power. 

Ensuring optimal resource allocation in fogs is crucial to efficiently handle dynamic IIoT workloads---with the minimum cost and delay. In another research \cite{kumar2023autonomic}, Kumar \etal have introduced a framework that presents an efficient approach for workload prediction and optimal placement of fog nodes to execute dynamic IIoT workloads. The approach incorporates a Deep Auto Encoder (DAE) model to forecast workloads. Then, based on the demand for IIoT workloads, the fog nodes are scaled. To optimize multiple objectives and meet service constraints, a meta-heuristic algorithm is developed. While these work considers a hierarchical federation of computing platforms across edge-to-cloud and execution cost as an important objective, our research considers a horizontal federation using peer computing platforms (\ie fog systems) which is more viable in remote industrial areas.


\subsection{Fog Computing for Industry 4.0 Use Cases}
In traditional operational systems of the industrial sector, the centralized cloud can only support legacy applications (\cite{alam2021cloud,shastry2022approaches}) with a considerable latency-tolerant nature. On the other hand, Industry 4.0 applications are latency-sensitive and need a dynamic execution platform to support (near) real-time response time. Fog computing has shown great potential in handling time-sensitive tasks for the Industrial Internet of Things (IIoT). However, resource allocation for fog nodes presents difficulties due to their limited capacity. To tackle this issue, Kumar \etal in \cite{kumar2023ai} develop a framework using artificial intelligence (AI). This system features a fuzzy-based offloading controller and an AI-powered Whale Optimization Algorithm (WOA) to enhance the Quality-of-Service (QoS) parameters. The primary aim of this research is to conserve the battery life of fog devices while our work ensures the successful completion of micro-services on fog nodes within strict time constraints. In a similar research, Rao \etal in (\cite{rao2021eco}) proposed a dynamic runtime for smart industrial applications that utilize 5G technology with edge-cloud architecture. This work uses application-specific knowledge to map the micro-services into the execution platform. However, authors consider only the predefined critical path's latency and disregard other micro-services that could form a new the critical path if they are poorly allocated. Additionally, they consider utilizing cloud data centers to overcome the emergency and oversubscribed situations. In contrast, our proposed solution considers all the micro-services (including those not on a critical path) to complete the execution across the fog federation with respect to the application deadline. Similarly, Faticanti \etal (\cite{faticanti2020throughput}) analyze the throughput desires of the micro-service applications while perform offloading to other fog systems. They addressed resource allocation challenges for the fog-native applications, built on a containerized micro-service modules. Two cascading algorithms make up the entirety of their approach. The first one separates fog application components according to throughput, and the second one governs the application orchestration across geographically distributed data centers.  
\subsection{ML Applications and Serverless Platform}
As smart IoT-based solutions become widespread, at the application level, there is a need to support ML applications on the network edge. However, such ML solutions are yet to be applicable at the fog federation platform level. This is mainly because existing ML-based resource allocation methods (\eg \cite{kumar2022cooperative,salmani2023reconciling,ruiz2023performance}) are trained for homogeneous and fairly stable underlying fog systems, whereas, in fog federation we encounter a truly heterogeneous distributed infrastructure that is overly dynamic. Moreover, although lightweight (edge-friendly) AutoML solutions (that employ ML for automation purposes (\cite{garouani2022amlbid})) operating based on model compression, pruning, and quantization is becoming a commonplace, there is yet to be pe-trained ML models for Industry 4.0 use cases deployed under the federated fog system dynamism. 

A related research work (\cite{ishakian2018serving}) examines the suitability of using a serverless architecture for AI application workloads. The research work evaluates the performance of using serverless functions to classify images using deep learning models. The authors find that warm serverless function executions have acceptable latency. However, cold starts have significant overhead, which could pose a problem for adherence to SLAs that do not account for this bimodal latency distribution. As industrial emergencies are unpredictable and rare, the system admin might not consider keeping active (a.k.a warm start) the resource orchestration functions due to high expense of execution. Furthermore, serverless frameworks lack access to GPUs, and functions are stateless, meaning each execution can only use CPU resources and cannot rely on the serverless platform runtime to keep state between invocations to optimize performance. As a result, serverless solutions will need to accommodate more stateful workloads in the long run. To achieve performance comparable to current non-serverless platforms, a declarative way of describing workloads and their requirements will be necessary, such as specifying the minimum time to keep warm containers (\cite{kumari2022mitigating}) and access to GPUs (\cite{risco2021gpu}). In an associated field of study (\cite{zhang2021serverless}), Zhang \etal explore and propose a serverless framework for video analytics pipelines based on deep neural networks (DNNs), which fully leverages the collaborative potential of client-fog-cloud architecture. By effectively coordinating fog and cloud resources, they achieve cost-effective and highly accurate video analytics. The proposed system incorporates a human-in-the-loop design approach to continuously improve the performance of models. Moreover, it offers various functionalities for creating and deploying video applications, alleviating developers from time-consuming tasks related to resource management and system administration. However, this work is curated for video processing tasks and incorporates cloud computing that can generate latency in time-sensitive smart applications.

\subsection{Software-hardware System Stack}
Patterson \etal (\cite{patterson2021hardware}) have addressed the need for having a specialized software-hardware system stack and a coordination mechanism to support edge-based applications. They propose a swarm coordination platform for the efficient and scalable execution of workflows of complicated tasks using cloud and edge resources. Their hardware-software system stack tries to bridge the gap between centralized and distributed coordination. Their suggested approach uses domain-specific languages (DSL) to enhance programmability, automatically mapping tasks to cloud and edge resources, and providing hardware acceleration fabrics for remote memory access and networking. However, they do not consider oversubscribed situations and latency constraint smart applications. In contrast, our work considers oversubscribed situation in remote areas where monolithic and micro-service applications coexist. 

\section{System Model}\label{sysModel}
\begin{figure}[h!]
	\centering	
	\includegraphics[width=0.50\textwidth]{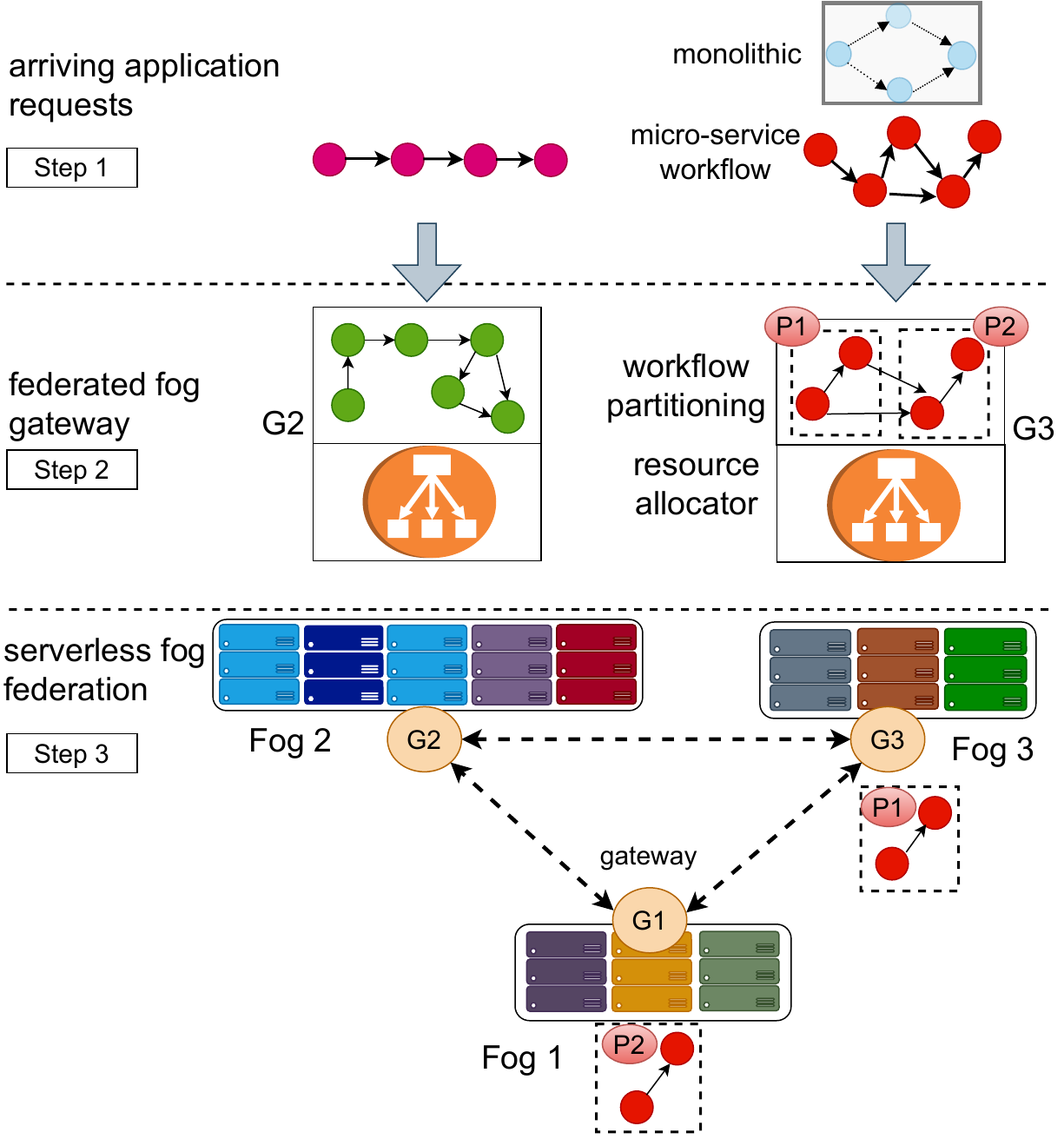}
	\caption{System model of the proposed solution to allocate micro-service workflow or monolithic applications across a federation of three fog systems. Step 1 shows different applications arriving to the gateway of their local fog system. Step 2 shows the internal mechanics of each gateway that includes a ``graph partitioning'' and a ``resource allocator'' modules. The former is in charge of transparently partitioning the micro-service workflows to maximize its likelihood of on-time completion. The latter is in charge of seamlessly allocating each partition to a fog across the federation. Step 3 shows the serverless fog federation where each fog is represented by a gateway ($G_1, G_2, G_3$).\label{fig:sysModel}}
\end{figure}

Industry 4.0 applications are often in form of micro-service workflows deployed via containers (\cite{pallewatta2023placement,roda2023cloud,wu2023towards}). As shown in Figure~\ref{fig:sysModel}, the system model of the proposed serverless fog federation comprises: (1) A gateway in each fog system that receives arriving application requests and transparently allocates them across the federation by considering the communications and computing uncertainties, the application architecture (\ie monolithic, micro-service), and its latency requirements. (2) A network of wireless gateways where each gateway $G_i$ represents a fog system that takes care of the coordination and data transfer with peer fogs across the federation. Each fog is an autonomous entity with its own resources and policies. 

The fog-to-fog communication is often based on Device-to-device (D2D) wireless communication that depends on various parameters, such as position in the federation, mobility, required coverage, data volume, environment, and spectrum to utilize (\cite{ogundoyin2022secure}). Considering the geographical area coverage required for the fog federation, Wireless Wan (WWAN) technologies, particularly 5G, are recommended by the OpenFog consortium \cite{openfog2017openfog}. 
5G can facilitate low-latency communication between fog systems (\cite{attaran2023impact}) and deliver peak data transmission speeds up to 10 Gbps (\cite{lagorio20235g}).

In this research, we consider an oversubscribed situation in a fog that is defined as a situation where the system cannot complete all the arriving requests within their deadlines. Each workflow application, denoted $\omega$, is a $DAG = (V,E)$, where the set of vertices $V = \{m_1, m_2,..., m_n \}$ denotes the micro-services, and edge $e(m_i,m_j) \in E$ represents the precedence between $m_i$ and $m_j$ micro-services. Each micro-service $m_i$ has a known slack (a.k.a. deadline and denoted $m_i^\delta$), which is the time duration within that the micro-service has to complete its execution. Furthermore, for each micro-service $m_i$, we assume to know the statistical distribution of its computational latency (denoted $m_i^d$) representing the possible execution times of that micro-service on that fog. Such a statistical distribution can be obtained from profiling the past execution times of the same micro-service on the same fog system (\cite{salehi2016stochastic,gentry2019robust}). 

Monolithic applications have ``full cohesion", thus, have a single deadline for the entire application. While partitioning in monolithic applications is infeasible, micro-service-based workflows have a ``loose cohesion'' and can be partitioned at the micro-service granularity (see Step 1 in Figure~\ref{fig:sysModel}). We designate the resource allocation method to operate after partitioning a micro-service workflow to allocate its partitions (containerized services) across the fog federation. As shown in Step 2 of the same figure, for each arriving request for a certain application, the fog gateway is responsible of creating the serverless illusion for the users by taking care of two actions: (\textbf{a}) optimally partitioning the workflow (\eg partitions \circled{P1} and \circled{P2} in Figure~\ref{fig:sysModel}); and (\textbf{b}) allocating resources to each partition via fog gateways (\eg $G1$, $G2$, and $G3$ in Figure~\ref{fig:sysModel}) across the federation. Finally, in Step 3, the serverless fog federation executes the applications and generates the results. 

We assume the types of industrial applications operating in the system are limited and known in advance. Accordingly, the gateway stores and updates the computation and communication latencies of various categories of applications on the serverless fog systems. The computational latencies for various micro-services are captured in a matrix, termed as Estimated Task Completion (ETC) where each entry represents computational latencies for a particular micro-service across the fogs in the federation. Similarly, the communication latencies for various micro-services reaching various fog systems of the federation is stored in a matrix table and named as Estimated Task Transfer (ETT). Hence, we capture the time for transferring various types of micro-services across fog federation under various circumstances (\ie different network congestion and  data dependencies). Therefore, all the communication and networking overheads are abstracted in the ETT entries. The information from these two matrices are employed to estimate the applications' time constraints. That is, the gateway relies on this information to make informed decisions regarding resource allocation across the fog federation.


\section{Partitioning Micro-service Application Workflows}\label{partitioning}

\begin{figure}[h!]
	\centering	
	\includegraphics[width=0.48\textwidth]{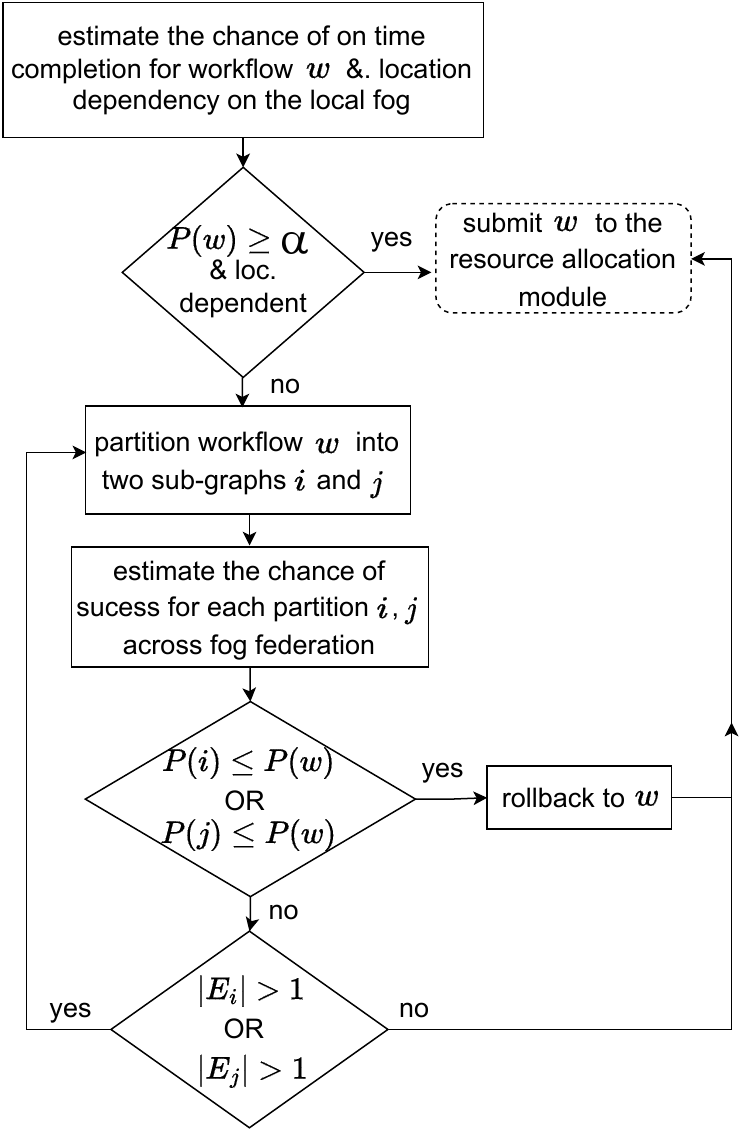}
	\caption{Flowchart of the ProPart workflow partitioning method. The output of this method is a partitioned workflow that is submitted to the resource allocation module, which is shown as the end step (with dashed lines) in this flowchart.\label{fig:partitionTech}}
\end{figure}  



The main objective of the partitioning method is to partition a micro-service workflow application in a way that the application can meet its deadline. For that purpose, we need to measure the likelihood of on-time completion for workflow graph $\omega$, denoted $P(\omega)$, on a fog system. This requires knowing the probability of workflow completion before its deadline. To know the workflow deadline, we sum up the deadlines of its micro-services, \ie $ \delta_\omega = \Sigma_{i=1}^n m_i^\delta$. To learn the workflow completion, denoted $D_\omega$, we need to convolve the computational latency distributions of the micro-services in the critical path of $\omega$ that can be performed based on Equation~\ref{eq:convolve}. 
\begin{equation}\label{eq:convolve}
D_\omega = m_1^d \circledast m_2^d \circledast ....\circledast m_n^d    
\end{equation}

Using the completion time distribution ($D_\omega$), the probability of on-time completion for $\omega$ is the portion of $D_\omega$ that occurs before $\delta_\omega$ which is measured based on Equation~\ref{eq:prob}.

\begin{equation}\label{eq:prob}
P(\omega) =\mathbb P(D_\omega \leq \delta_\omega)
\end{equation}

Leveraging this theory, we can present our workflow partitioning method, called Probabilistic Partitioning (\textit{ProPart}). The flowchart of ProPart, shown in Figure~\ref{fig:partitionTech}, shows the steps to partition workflow $\omega$. In the first step, we assume executing the entire workflow on the receiving (local) fog system without any partitioning and calculate $P(\omega)$. The value of $P(\omega)$ is decisive on partitioning the workflow across the fog federation or executing it locally---on the receiving fog system---without partitioning it. Furthermore, we check for location dependency of the micro-services within the workflow on the local fog. To make the decision, the value of $P(\omega)$ is compared against a user-define value $\alpha$ (see Figure~\ref{fig:partitionTech}) that serves as a threshold determining how aggressively the user wants to take advantage of the federation. The lower values of $\alpha$ expresses the user tendency to execute $\omega$ locally and vice versa. In the extreme case, the user can forbid using the federation via choosing $\alpha=0$.
In our implementation, the value of $\alpha$ is adjustable, however, in the evaluation part, we consider $\alpha=50\%$. In addition, we check location dependency of the micro-services that is one of the prerequisite for successful completion of the workflow $w$ mentioned in earlier sections.

In the event that the likelihood of completing workflow $\omega$ is lower than $\alpha$, the graph partitioning operation has to be carried out. Finding the optimal number of partitions is an NP-complete problem (\cite{ccatalyurek2023more}), therefore, we develop a method based on the idea of divide and conquer (\cite{zhou2023qaoa}) to recursively find the most efficient way of partitioning $\omega$. The bare bones of the method is shown in the flowchart of Figure~\ref{fig:partitionTech}. At every step of this method, the graph is partitioned to two subgraphs, and for each subgraph the likelihood of success across the federation is calculated based on Equation~\ref{eq:prob}. The process continues until either the likelihood of success for at least one of the subgraphs is lower than or equal the parent graph, or the subgraphs have only one vertex (node). The former case implies that partitioning the graph does not improve the chance of success, hence, we roll-back to the parent graph and consider that subgraph as the appropriate partition that can be allocated to the fog federation. In the latter case, however, the graph cannot be partitioned any further and inevitably must be submitted for the resource allocator module.

To perform graph partitioning, we use the min-cut algorithm (\cite{lakhan2022deep}) that partitions workflow $\omega$ into two sub-graphs $i$ and $j$ based on the max-flow min-cut theorem (\cite{lochbihler2022mechanized}). The theorem states that the maximum flow (\eg amount of data) that can be transferred from a certain vertex (\eg micro-service) to another one is determined by the smallest bottleneck in the graph. Hence, to partition the workflow graph into sub-graphs, the min-cut algorithm finds the minimum number of edges that, if removed, the graph is partitioned to subgraphs. 

According to Flowchart of Figure~\ref{fig:partitionTech}, at each iteration of the method, the partitions (subgraphs) resulted from the min-cut are estimated for their chances of success (meeting their deadlines) across the fog federation using Equations~\ref{eq:convolve} and~\ref{eq:prob}. If the likelihood of on-time completion of the partitioned graph is less than its parent graph (\ie $P(i) \leq P(\omega)$ or $P(j) \leq P(\omega)$), we consider the parent graph ($\omega$) as the optimal choice for allocation. In the next step, for each created partition, the resource allocation method is invoked (the last box in the flowchart). In the event that the partition's chance of on-time completion is greater than its parent graph, then we examine another iteration of partitioning. The stopping condition for partitioning is to reach graphs with only one vertex (one micro-service).

\section{Resource Allocation Method across Serverless Fog Federation}\label{resourceAllocation} 
The resource allocation module is in charge of allocating arriving requests---either in form of monolithic applications or partitioned micro-service workflows---across the local or the federated fog systems. The allocation is performed based on the notion of \textit{relevance} that is defined as a fog system that maximizes the likelihood of meeting the deadline (a.k.a. the \emph{probability of success}) for a given request. The probability of success for a single unit of execution (\ie a micro-service or a monolithic application) on a particular fog system depends on the \textit{end-to-end latency} distribution of that unit on that fog.

The end-to-end latency is comprised of the communication and computation latencies.
Each fog system uses historical computational and communication latency information of various micro-services across different fogs in the federation to generate the probability distributed functions (PDFs). In analogy, this is very much like routing tables (\cite{fall2009routing}) commonly used by the Internet routers. Two matrices are maintained at each gateway, namely, Estimated Task Computation (ETC) (\cite{diaz2011energy}), and Estimated Task Transfer (ETT) (\cite{hussain2019federated}). 
The PDF of computational latency for micro-service type $i$ on fog system $j$ is stored in entry $ETC(i,j)$ that is also used by the partitioning method, as noted in the previous section. Similarly, entry $ETT(i,j)$ maintains the communication latency PDF of transferring data for micro-service $i$ to fog $j$. We note that using these matrices makes the resource allocation method aware of the communication latencies, whereas,  partitioning method is less granular and only is aware of the computational latency. There are existing methods based on machine learning or other statistical approaches (\eg \cite{,}) to construct and regularly update $ETC$ and $EET$ matrices without any interference on the system functionality.

For partition $\omega$ with source node (micro-service) $i$, the resource allocation method computes its end-to-end latency distribution on fog $j$ via convolving $D_\omega$ (see Equation~\ref{eq:convolve}) and $ETT(i,j)$. The end-to-end latency distribution is then used in Equation~\ref{eq:prob} to calculate the probability of completing partition $\omega$ before its deadline. Once we know the chance of success on all adjacent fogs in the federating, the one that offers the highest probability of on-time completion is chosen as the assignment destination for $\omega$.

It is noteworthy that the probability of completing $\omega$ on the fog originally receives it (a.k.a. local fog) does not include any communication latency. This means that, $\omega$ is assigned to an adjacent fog in the federation (a.k.a. remote fog), only if it offers a greater probability of on-time completion even after accounting for the communication latency. However, utilizing a remote fog in a remote industrial place implies environmental uncertainties, in addition to the communication and computation uncertainties, that can practically undermine the benefits of minor superiority in the probability of on-time completion. That is, the likelihood of on-time completion of $\omega$ remotely should be \textit{significantly} higher than the local one, so that it is worthwhile allocating it remotely.
To assess whether the difference between remote and local execution is significant, we propose to use the notion of confidence intervals (CI) of the underlying end-to-end completion time distributions. In particular, we check whether the CI of the end-to-end completion time distribution of the remote fog overlaps with the CI of the local one. If they do not overlap, it implies that the remote fog is offering a statistically and practically higher chance of success to $\omega$, otherwise it is not worthwhile allocating it remotely.

The pseudo-code provided in Algorithm~\ref{alg:ci} expresses the resource allocation method that each gateway utilizes to take advantage of the federated fog system. The method is called \emph{Maximum Probability (MR)} and is invoked for the set of partitioned micro-service workflows, denoted as $M$. 
Using the deadline of each micro-service partition $\omega \in M$, from Step 2---10 of the algorithm, the algorithm calculates the local and remote completion time distributions and its probability of success both locally and on the adjacent fog systems. 
Next, Step 11 sorts the calculated probabilities on remote fogs in the descending order. If the probability of success on the local fog is higher, then $\omega$ is assigned to the local fog. Otherwise, $\omega$ is assigned to a remote fog if the CI of its end-to-end latency distribution (denoted $E_g(\omega)$) does not overlap with the distribution of the local one (Step 14---17). Otherwise, the same procedure is performed for the rest of the neighboring fogs. If no non-overlap adjacent fog is found, then $\omega$ is assigned to the local fog (default assignment in Step 12).

\begin{algorithm}[!h]
	\SetAlgoLined\DontPrintSemicolon
	\SetKwInOut{Input}{Input}
	\SetKwInOut{Output}{Output}
	\SetKwFunction{algo}{algo}
	\SetKwFunction{proc}{Procedure}{}{}
	\SetKwFunction{main}{\textbf{TaskAssignment}}
	\Input{set of partitions $M$; $ETC$ and $ETT$ matrices; and set of the adjacent fogs $G$}
	\Output{chosen fog $g\in G$ to assign each partition $\omega \in M$}
        \ForEach{$\omega \in M$}{
        $E_r(\omega) \gets$ completion time distribution of $\omega$ on local fog\;
       $P_r(\omega) \gets$ probability of success on local fog \;
	\ForEach{$g \in G$} {
        $E_g(\omega) \gets$ end-to-end completion distribution of $\omega$ on fog $g$\;
	  $P_g(\omega) \gets$ probability of success on $g$ \;
          \If {$P_g(\omega) > P_r(\omega)$} {	
				      \small{add $P_g(\omega)$ to $F$, as a potential fog for assignment}\;
       	        	        }
       	}        	        
	sort $F$ in descending order\;
        consider local fog $r$ as default assignment for $\omega$ \;
	\ForEach{$P_g \in F$} {
	\If {CI $E_g(\omega)$ has no overlap with CI $E_r(\omega)$} {	
				      assign $\omega$ to fog $g$ \;
				      exit the loop\;   	
       	        	        }
       	}        	        
	}
\caption{Pseudo-code of the Maximum Probability (MR) resource allocation method}\label{alg:ci}
\end{algorithm}

\section{Performance Evaluation}\label{result}
This section evaluates the proposed partitioning and resource allocation methods against baseline solutions. Here, we describe the simulation setup and scenarios, define baseline methods, and present practical experiments with analysis.
\subsection{Experimental Setup}
EdgeCloudSim (\cite{sonmez2017edgecloudsim}) is a discrete event simulator that we have used to develop the solution and then evaluate its performance. We simulate fog systems with eight processing nodes that process between 1500 and 2500 Million Instructions Per Second (MIPs) that represent the heterogeneity of fog systems across the federation. The synthesized fog federation is considered to be distributed across a two-dimensional plane. Hence, every fog has coordinates defined by the x and y-axis. Accordingly, each fog can have a maximum of four neighbours, and the federation is horizontally scalable.

\begin{table}[h!]
\centering
\scalebox{0.70}
{\begin{tabular}{l||l|l|l|l}

\textbf{Application} & \textbf{DNN Model} & \textbf{Input Type} & \textbf{Scope} & \textbf{Framework}\\ \hline \hline
\textit{Fire} & Customized Alexnet & Video Segment & \begin{tabular}[c]{@{}l@{}}Control \&\\ Monitoring\end{tabular} & \begin{tabular}[c]{@{}l@{}}Tensorflow \\ (tflearn)\end{tabular}\\ \hline
\textit{\begin{tabular}[c]{@{}l@{}}HAR\end{tabular}} & \begin{tabular}[c]{@{}l@{}}Customized Sequential \\ Neural Network\end{tabular} & Motion sensors & \begin{tabular}[c]{@{}l@{}}Workers \\ Safety\end{tabular} & keras\\ \hline
\textit{Oil} & FCN-8 & SAR Images & \begin{tabular}[c]{@{}l@{}}Disaster \\ Management\end{tabular} & keras\\ \hline
\textit{\begin{tabular}[c]{@{}l@{}}AIE \end{tabular}} & \begin{tabular}[c]{@{}l@{}}Temporal Convolutional\\ Network\end{tabular} & Seismic Data & \begin{tabular}[c]{@{}l@{}}Seismic\\ Exploration\end{tabular} & PyTorch\\ 
\end{tabular}}
\caption{\small{ML-based applications used in many Industry 4.0 use cases, such as oil and gas, along with their neural network model, input data type, usage scope, and ML framework. The non-abbreviated application names in this table are: fire detection (Fire), human activity recognition (HAR), oil spill detection (Oil), and acoustic impedance estimation (AIE).}}
\label{table:apptype}
\end{table}

\begin{table}[h!]
\centering
\scalebox{0.75}
{\begin{tabular}{|c|c|c|c|c|c|}
\hline
\multicolumn{6}{|c|}{\textbf{Mean and Standard Deviation of Execution Times (ms) on AWS}} \\ \hline
\textbf{Application} & \textbf{\texttt{Mem. Opt.}} & \textbf{\texttt{ML Opt.}} & \textbf{\texttt{GPU}} & \textbf{\texttt{General}} & \textbf{\texttt{Compute Opt.}} \\ \hline
\textit{Fire} & \begin{tabular}[c]{@{}c@{}}$\mu$=1461.8\\ $\sigma$=457.3\end{tabular} & \begin{tabular}[c]{@{}c@{}}$\mu$=1281.7\\ $\sigma$=387.93\end{tabular} & \begin{tabular}[c]{@{}c@{}}$\mu$=1349.5\\ $\sigma$=418.9\end{tabular} & \begin{tabular}[c]{@{}c@{}}$\mu$ =1534.8\\ $\sigma$=494.7\end{tabular} & \begin{tabular}[c]{@{}c@{}}$\mu$=1421.4\\ $\sigma$=441.8\end{tabular} \\ \hline
\textit{\begin{tabular}[c]{@{}c@{}}HAR\end{tabular}} & \begin{tabular}[c]{@{}c@{}}$\mu$=1.27\\ $\sigma$=0.082\end{tabular} & \begin{tabular}[c]{@{}c@{}}$\mu$=0.66\\ $\sigma$=0.006\end{tabular} & \begin{tabular}[c]{@{}c@{}}$\mu$=0.51\\ $\sigma$=0.006\end{tabular} & \begin{tabular}[c]{@{}c@{}}$\mu$ =1.17\\ $\sigma$=0.042\end{tabular} & \begin{tabular}[c]{@{}c@{}}$\mu$=0.66\\ $\sigma$=0.003\end{tabular} \\ \hline
\textit{\begin{tabular}[c]{@{}c@{}}Oil\end{tabular}} & \begin{tabular}[c]{@{}c@{}}$\mu$=269.9\\ $\sigma$=1.01\end{tabular} & \begin{tabular}[c]{@{}c@{}}$\mu$=218.8\\ $\sigma$=0.66\end{tabular} & \begin{tabular}[c]{@{}c@{}}$\mu$=65.98\\ $\sigma$=0.47\end{tabular} & \begin{tabular}[c]{@{}c@{}}$\mu$=667.1\\ $\sigma$=2.26\end{tabular} & \begin{tabular}[c]{@{}c@{}}$\mu$=242.9\\ $\sigma$=0.68\end{tabular} \\ \hline
\textit{\begin{tabular}[c]{@{}c@{}}AIE\end{tabular}} & \begin{tabular}[c]{@{}c@{}}$\mu$=7.02\\ $\sigma$=0.02\end{tabular} & \begin{tabular}[c]{@{}c@{}}$\mu$=6.41\\ $\sigma$=0.03\end{tabular} & \begin{tabular}[c]{@{}c@{}}$\mu$=7.55\\ $\sigma$=0.04\end{tabular} & \begin{tabular}[c]{@{}c@{}}$\mu$=9.35\\ $\sigma$=0.06\end{tabular} & \begin{tabular}[c]{@{}c@{}}$\mu$=7.95\\ $\sigma$=0.02\end{tabular} \\ \hline
\end{tabular}}
\caption{\small{The execution time mean ($\mu$), and standard deviation ($\sigma$) for various Industry 4.0 applications across different AWS machines types.}}
\label{tab:meanAWS}
\end{table}

As for the workload to evaluate our methods, we consider the context of remote Industry 4.0 for oil and gas (\cite{hussain2022iot}). We consider four ML-based workflow applications, namely, Fire detection ( \textit{Fire}), human activity recognition (\textit{HAR}), oil spill detection (\textit{Oil}), and acoustic impedance estimation (\textit{AIE}) (\cite{hussain2020analyzing}) and they are uniformly distributed within the workload. Hence, total number of applications in the workload are equally divided into the four Industry 4.0 applications.

While exploring the micro-service workflow architecture, as shown in Figure \ref{fig:workflowMicro}, we find that there are seven micro-services for fire detection application. Similarly, the number of micro-services for oil spill detection, human activity recognition, acoustic impedance estimation are five, four, and four sequential (single-core) micro-services, respectively. In Oil, the micro-services forming the workflow are data pre-processing, dark spot detection, feature extraction, classification, and segmentation micro-services. For AIE application, the micro-services are for data pre-processing, initial model develop, inversion, and acoustic impedance estimation. For HAR application we use data pre-processing, feature extraction, classification, and activity recognition micro-services. Other characteristics of these applications are presented in Table \ref{table:apptype}. We benchmarked these applications on AWS and identified their statistical distributions of their execution times (in terms of MIPS) that are presented in Table~\ref{tab:meanAWS}. 
For monolithic workload, we use the same set of applications, however, we treat them as one unit and do not partition them. One reason that we study monolithic applications in the experiment section is to abstract our analysis from the impact of workflows, and merely focus on the impact of the resource allocation methods across the federation.

Upon arrival of a request to a gateway, the micro-services comprising the application are allocated distinct deadlines, as mentioned in our system model. According to reference (\cite{salehi2016stochastic}), an individual deadline comprises the time a micro-service arrives and the total delay that the micro-service can withstand in an end-to-end scenario. The delay in communication, which includes both up-link and down-link delays, can substantially affect the deadline. Accordingly, we consider the communication delay in the deadline calculation. For micro-service $i$, the deadline $m_i^d$ is defined as $m_i^d$ = $arr_i$+$E_i$+$\epsilon$+$d_C$ , where $arr_i$ is the arrival time of request, $E_i$ is the average micro-service execution time, $\epsilon$ is a constant value defined by the processing fog device (slack time), and $d_c$ is the mean communication delay. To produce the inter-arrival rate of the requests, we synthesized from real-world workload investigated by the Extreme Scale Systems Center (ESSC) at Oak Ridge National Laboratory (ORNL) (\cite{khemka2014utility,khemka2015utility}). In order to remove the impact of any randomness in the evaluations, we carry out every experiment a total of 30 times and subsequently estimate the mean and 95\% confidence interval. 

As we work towards streamlined and efficient solutions for Industry 4.0, we evaluate suggested approaches alongside widely used fundamental statistical computations rather than relying on intricate ML-based network models. Given that our focus is on the successful execution of ML-based micro-service applications, we are steering clear of system-level intricacy by forgoing complex ML-based models, which may necessitate sophisticated hardware support and real-time data to achieve optimal network model training.


\subsection{Baseline Workflow Partitioning Methods}
\emph{Min-Cut partitioning:} cutting graph $G$ partitions its vertices (micro-services) into two disjoint proper subsets. In a weighted graph, the cost of a cut is the sum of weights that are involved in the cut. Hence, the term ``minimum cut'' refers to a cut that is either minimal in terms of the number of edges that cross the cut (when the edges are not weighted) or minimal in terms of the weights of the edges that cross the cut (weighted). Min-cut is a widely-used method in the literature (\cite{lochbihler2022mechanized}), therefore, we use it as a baseline method via considering a unit weight for edges of a workflow graph. 

\emph{Least data transfer:} Within a workflow, output of a preceding micro-service serves as the input data for another micro-service. The ``Least data transfer'' partitioning method (\cite{ahmad2014data}) considers the magnitude of data-transfer between micro-services as the weights of edges. Accordingly, this method traverse all the edges and cuts the workflow from the point where the transferred data is minimum. 

\subsection{Baseline Resource Allocation Methods}
\emph{Minimum Expected Completion Time (MECT):}
This is a popular resource allocation method~(\cite{mokhtari2023e2c,salehi2016stochastic}) that, for each request arriving to the gateway, it utilizes the ETC matrix to compute the mean expected completion time across different fog systems. The fog system with the minimum expected completion time is then chosen to handle the request.

\emph{Maximum Computation Certainty (MCC):}
MCC is another method used in the literature (\eg \cite{hussain2018robust}) for resource allocation. It makes use of the ETC matrix to calculate the difference between the request’s deadline and its mean completion time
(called certainty). Then, the task is assigned to the fog system that offers the highest certainty.

\subsection{Experimental Results}
The focus of this research is to explore the dynamic fog federation functionality upon facing a surge in demand for micro-service workflow applications with deadline constraints. Efficacy of the federation system in meeting the deadlines of these applications reflects its fault-tolerance. 

\subsubsection{Evaluating the Impact of Workflow Partitioning}
\begin{figure}[h!]
	\centering	
	\includegraphics[width=0.48\textwidth]{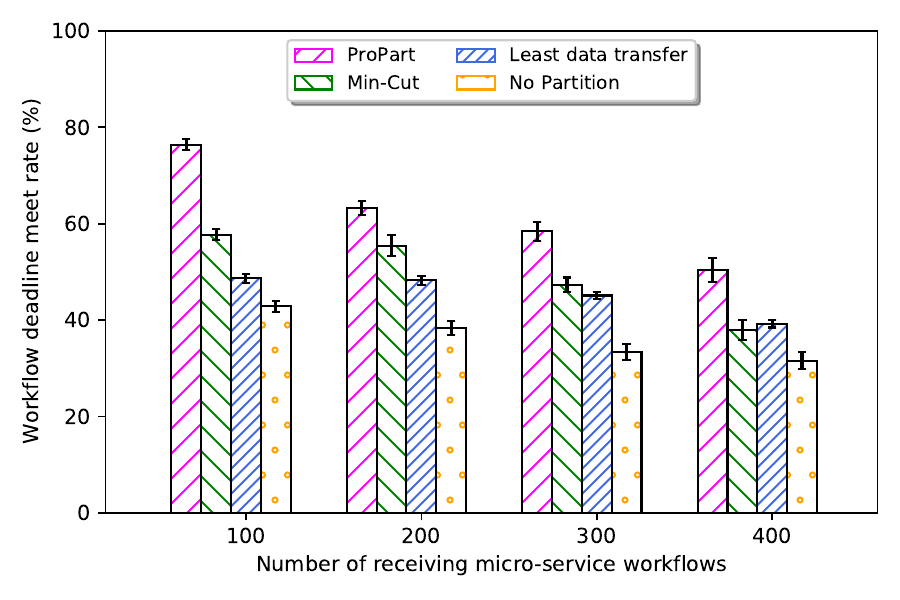}
	\caption{Comparison of the workflow partitioning methods. The horizontal axis represents the number of arriving workflow requests and the vertical axis shows the requests' deadline meeting rate.\label{fig:partitionComparison}}
\end{figure}

To examine the impact of workflow partitioning, in this experiment, we study a scenario where no method for workflow partitioning is in place, therefore, the whole workflow is either executed locally or outsourced to another fog in the federation. We compare this scenario against cases where partitioning can be performed when needed using ProPart method. To further study the impact of partitioning, we compare the performance (ProPart) against other baseline methods for partitioning, namely Min-cut and Least data transfer. For the purpose of this experiment, we increase the number of workflow requests submitted to a gateway to generate oversubscribed conditions and record the rate of meeting deadline for each case. For this experiment, we configure Max Probability (MR) as the resource allocation method. The result of this experiment is shown in Figure \ref{fig:partitionComparison}. The horizontal axis indicates the number of workflow requests received and the vertical axis shows the rate at which application deadlines have been met.

The experiment results indicate that, in all cases, the percentage of workflows meeting their deadlines decline upon increase in the number of requests. Nonetheless, the advantages of workflow partitioning is significant, with ``No partition'' invariably performing poorly. We witness that partitioning can improve the performance of using federation by at least 5\% (``No Partition'' versus ``Least data transfer'' with 400 requests), and at most 35\% (``No Partition'' versus ``ProPart'' with 100 requests). Moreover, across various partitioning methods, we observe that ProPart surpasses other partitioning methods (particularly when the system is not oversubscribed), whereas, Least data transfer yields the lowest deadline meet rate. The reason for the higher performance of ProPart is that, under lower load, its statistical analysis to obtain the likelihood of success leads to more accurate results, thereby, more informed partitioning decisions. Moreover, we observe that, as the number of requests rises, the performance of Min-cut gradually drops to the extent that for 400 requests, it even gets marginally worse than the Least data transfer method. The reason is that, Least data transfer partitions based on the edges that generate the smallest output data that is effective when the system becomes oversubscribed, whereas, Min-cut utilizes the least number of edges for partitioning and does not consider the amount of data travels between the partitions. 

\subsubsection{Evaluating the Impact of Resource Allocation}
\begin{figure}[h!]
    \centering
    \includegraphics[width=0.48\textwidth]{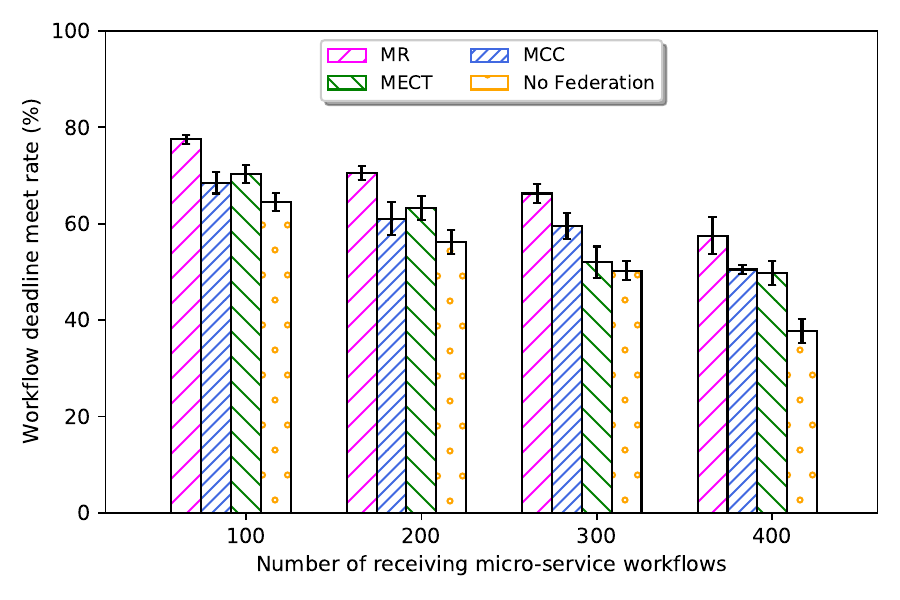}
    \caption{The deadline meet rates resulted from the MR, MCC, and MECT resource allocation methods of micro-service workflows across the fog federation.}
    \label{fig:resourcAllocationComp}
\end{figure}

\begin{figure}[h!]
    \centering
    \includegraphics[width=0.48\textwidth]{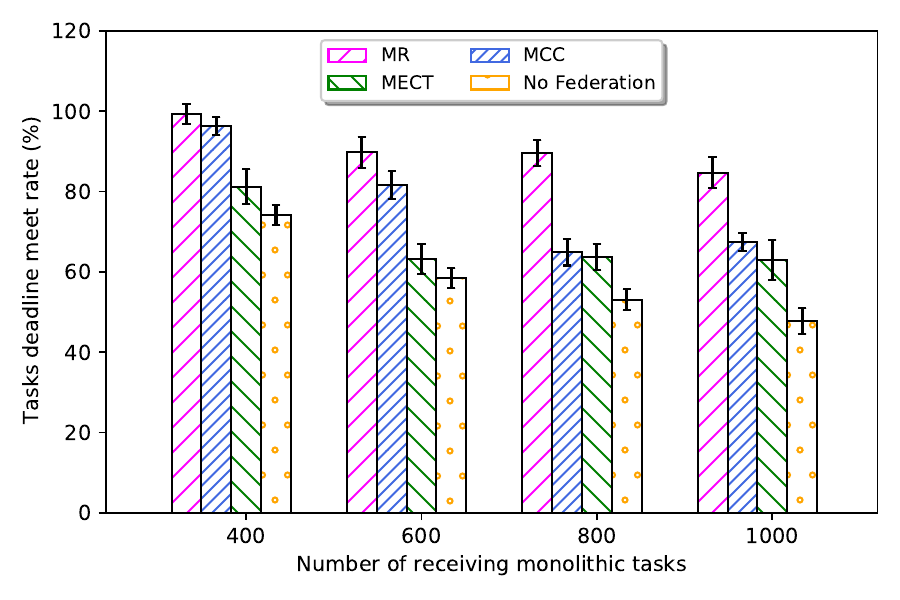}
    \caption{The deadline meet rate resulted from various resource allocation methods for monolithic applications across the fog federation.}
    \label{fig:resourcAllocationComp2}
\end{figure}

\begin{figure}[h!]
    \centering
    \includegraphics[width=0.48\textwidth]{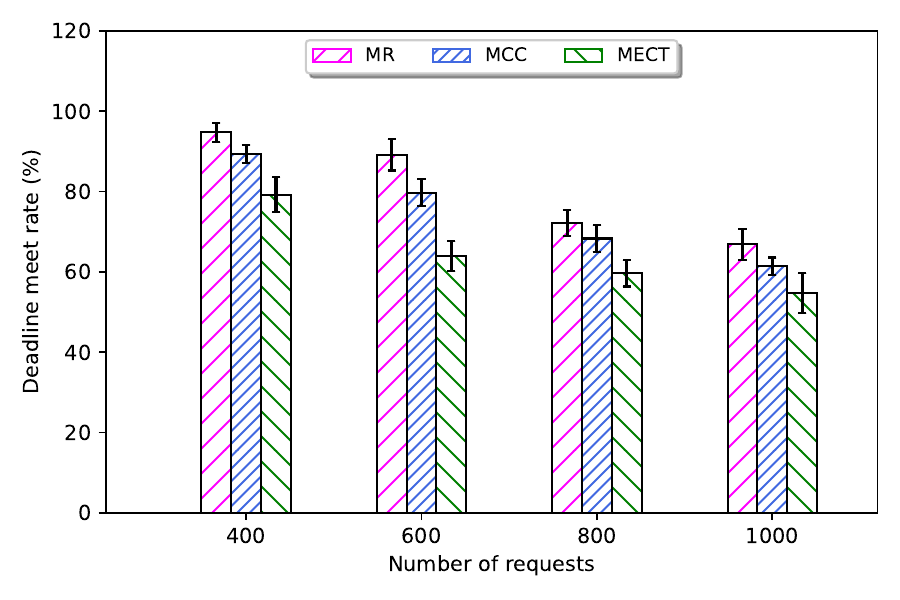}
    \caption{The deadline meet rate resulted from various resource allocation methods for mixed workload of monolithic and micro-service workflows across the fog federation.}
    \label{fig:mixWorkload}
\end{figure}

To study the impact of the resource allocation method in the fog federation, we performed the following experiments with three different resource allocation methods that, include the proposed MR method and two baseline methods, MECT and MCC. To studying the impact of using serverless federation fog versus not using it, we also include a scenario where no federation is utilized. We conducted this experiment for micro-service workflows, monolithic applications, mix application workload and the results are shown in Figures~\ref{fig:resourcAllocationComp} \ref{fig:resourcAllocationComp2}, and \ref{fig:mixWorkload} respectively.

\emph{Micro-service Workflow Applications:}  Similar to the previous experiment, the number of workflow requests is incremented to examine various levels of oversubscription, which is shown in the horizontal axis of Figure~\ref{fig:resourcAllocationComp}. In this experiment, we use ProPart to partition the micro-service applications. This implies that the MR bars, in the figure, show the performance of the entire methods we proposed in this work. To visualize the performance of the resource allocation methods, the deadline meet rates of receiving requests are reported. We can see that the benefits of making resource allocator aware of the fog federation is significant, because no federation scenario performed worse than the ``No Federation'' at any oversubscription level. This is because in a no-federation scenario, the recipient fog becomes overwhelmed with numerous requests, leading to missing request deadlines. Upon increasing oversubscription level, we can see a downward trend for all the resource allocation methods. However, because the proposed resource allocation method, MR, is aware of the compound uncertainty in both computation and communication, it still outperforms other baseline methods. In contrast, MECT and MCC are only cognizant of computation latency and the deadlines, therefore, they are prone to less informed allocation decisions that can cause missing the requests' deadlines. We can also observe that the performance difference of MCC and MECT is not statistically significant, because their confidence intervals overlap throughout the experiment. 

\emph{Monolithic Applications: } 
In this part, we investigate the performance of our resource allocation methods on the monolithic applications submitted to the system of fog federation. In addition we include a scenario where no federation is utilized to understand the benefits of using federated fog systems. In Figure \ref{fig:resourcAllocationComp2},  we can see the impact of increasing the number of incoming applications from 400 to 1,000 (horizontal axis) on the deadline meet rate (vertical axis) when various resource allocation heuristics are employed.
In fact, we performed the experiment for oversubscription levels less than 400, however, we did not observe any behavioral difference. As such, for the sake of better presentation, we only report results for cases where the number of requests are greater than 400.

Similar to the previous case, in Figure \ref{fig:resourcAllocationComp2}, we observe the poor performance resulted by ignoring the federation. The issue becomes increasingly pronounced as oversubscription level rises (800 and 1,000 requests), highlighting the advantage of serverless fog federation for Industrial use cases in remote areas. With 1,000 requests, MR offers around 19\% higher deadline-meeting rates than the other two methods. The reason is that MR captures end-to-end latency and utilizes the federation only if it significantly impacts the chance of success. From these experiments, we can conclude that for both monolithic and workflow applications, capturing uncertainty of the end-to-end latency can significantly improve the performance. This impact is more remarkable for monolithic tasks, \ie monolithic tasks can make a better use of the federation, because they do not have the challenge of request partitioning and performing one request across multiple sites.

\emph{Mixed Application Workload:} To further evaluate the impact of resource allocation, in this experiment, we consider circumstances where the arriving workload is a mix of requests for monolithic and workflow applications. 50\% of the examined workload are requests for monolithic applications and another 50\% are micro-service workflows. Similar to the previous parts, we increase the number of requests (horizontal axis in Figure \ref{fig:mixWorkload}) and measure the deadline meet rate (vertical axis in the same figure). We use ProPart for partitioning the workflow applications along with various resource allocations methods.

Decreasing the performance (deadline meet) upon increasing the number of requests aligns with our observations in the previous parts. However, we can see that even under mixed workload the combination of ProPart+MR consistently outperforms other methods for any level of request intensity. Although the performance difference of MR and other methods is not as significant as those in Figure~\ref{fig:resourcAllocationComp2} (for monolithic applications), it still outperforms the other methods due to the consideration of the end-to-end latency in the serverless fog federation. Analyzing other outputs of this experiment showed that, for monolithic requests, MR helped effective use of the federation and improving their performance, whereas, for micro-service workflow requests, the partitioning method contributes more than the resource allocation to the better performance.

\subsubsection{Analyzing the Workflow Makespan Time}
After scrutinizing behavior of the partitioning and resource allocation modules, in this part, our objective is to study the impact of entirety of the proposed solutions on the completion of requests. In fact, partitioning and and resource allocation methods can affect the computational latencies of the micro-service workflows. For the monolithic applications, resource allocation method is the influential factor on the requests' computational latencies. We measure the computational latency via the makespan time, which is the total processing time from the time a request is received until it is complete. Therefore, we capture the makespan time both for the micro-service and monolithic applications. We note that the experiments in the previous part only show the deadline meet rate and do not show how long on average all requests take to complete. 

\begin{figure}[h!]
    \centering
    \includegraphics[width=0.48\textwidth]
    {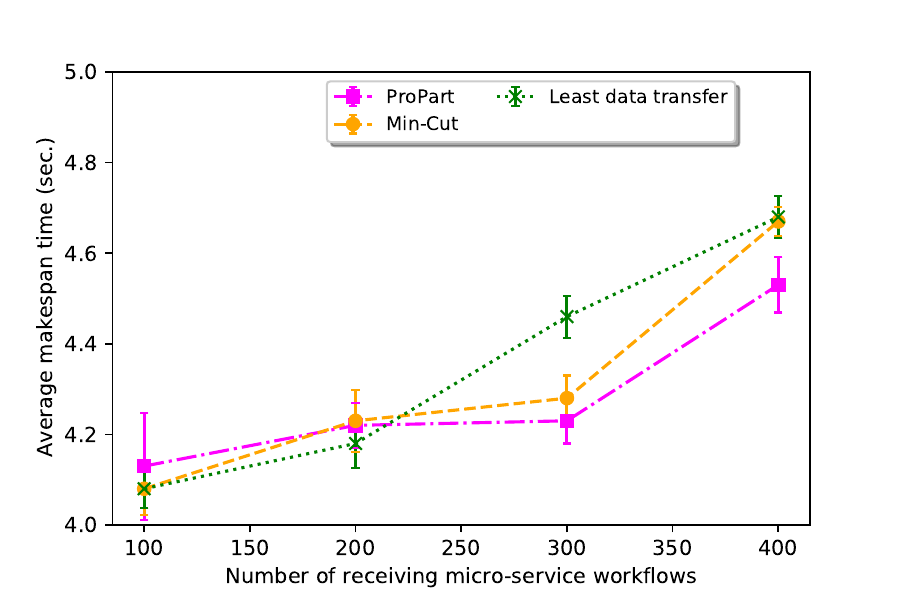}
    \caption{Average makespan time for micro-service workflows utilizing various partitioning methods along with the MR resource allocation method}
   \label{fig:microMakespan}
\end{figure}

\emph{Average Makespan Time for Micro-service Workflows:}
In this part, we estimate the average makespan time of the workflow requests on the fog federation. Similar to the previous experiments, we increase the oversubscribing level by submitting more workflow requests. The result of this experiment is presented as a line chart in Figure~\ref{fig:microMakespan} where the x-axis represents the number of receiving requests, and the y-axis represents the average makespan time. We examine various partitioning methods and keep MR, as the resource allocation method. 

The general observation shows an upward trend in the average makespan time, regardless of the partitioning method. The difference across methods become visible after 200 requests. We can see that ProPart outperforms the other two baselines upon increasing the oversubscription level. This is because ProPart performs partitioning only if the success rate of completion is significant. On the other hand, the baseline methods partition requests without considering the latency constraints. 

\emph{Average Makespan Time for Monolithic Applications: }
Due to cohesion of monolithic applications, their computational latency is only influenced by the resource allocation method. Hence, we study the impact of various resource allocation methods on the makespan of the monolithic applications across the fog federation.

\begin{figure}[h!]
    \centering
    \includegraphics[width=0.48\textwidth]
    {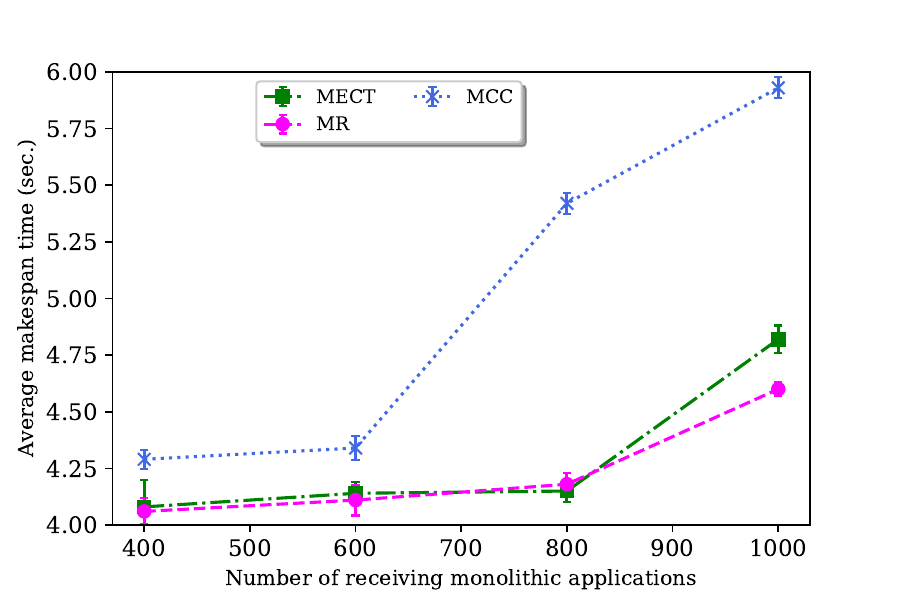}
    \caption{Average makespan time resulted from various resource allocation methods for the monolithic applications}
    \label{fig:monoMakespan}
\end{figure}

The result of this experiment is presented in Figure~\ref{fig:monoMakespan}, demonstrating that MR and MECT perform similarly unless the system is highly oversubscribed. Moreover, MCC performs significantly worse for any workload size. The reason is that MCC do not consider the stochastic nature of completion time, hence, it assigns the arriving requests to one fog and oversubscribes that, which increases the average makespan. From the experiment, we can conclude that if the system is not oversubscribed, the serverless fog federation system does not need to undergo the complication of methods like MR and other methods like MECT and even MCC can provide a competitive performance. However, if the system is expected to receive higher amount of load, then methods that consider factors such as end-to-end latency are needed to better cope with the oversubscription.

\subsubsection{The Impact of Fog Federation Scaling on Fault Intolerance}
The serverless fog federation in remote Industry 4.0 sites can potentially have a dynamic topology that has significant impact on fault intolerance (deadline meet rate) characteristic of the system. For instance, when a disaster occurs in an offshore oil field, one or more rescue boats with mounted fog systems can be dispatched to seamlessly augment the existing fog and help in handling the disaster (\cite{gima2019model}).This improves fault intolerance performance of the federation. Conversely, in the event that one or more of these mobile fogs fail or leave the disaster place, a performance drop in fault intolerance can potentially happen. The purpose of this experiment is to understand the impact of such dynamics on the overall fog federation's fault intolerance performance. 

\begin{figure}[h!]
    \centering
    \includegraphics[width=0.48\textwidth]{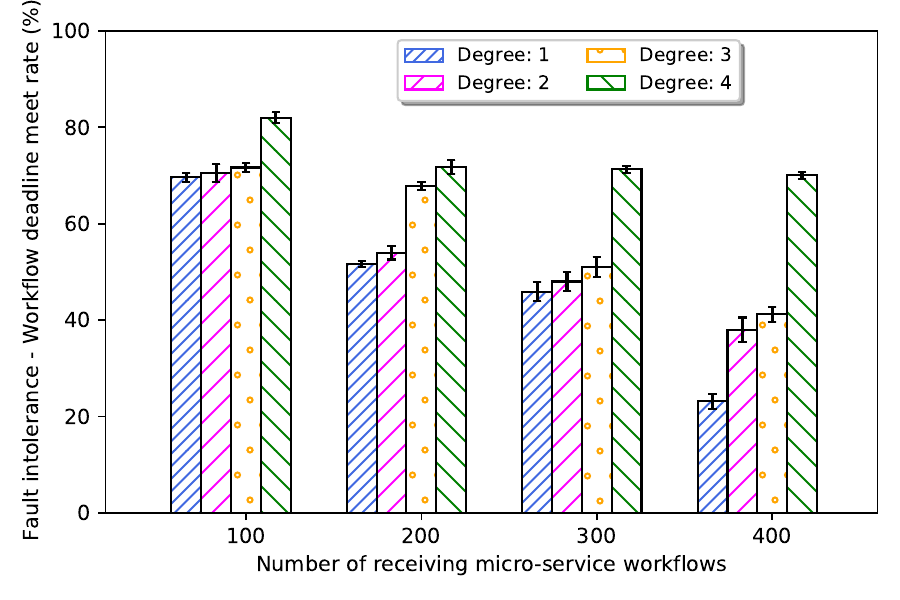}
    \caption{Impact of scaling the serverless fog federation using proposed partitioning and resource allocation methods upon increasing the oversubscription level for micro-service workflows against fault intolerance performance. The degree represents the number of neighbors each fog system has.}
    \label{fig:fedScaling}
\end{figure}

\begin{figure}[h!]
    \centering
    \includegraphics[width=0.48\textwidth]{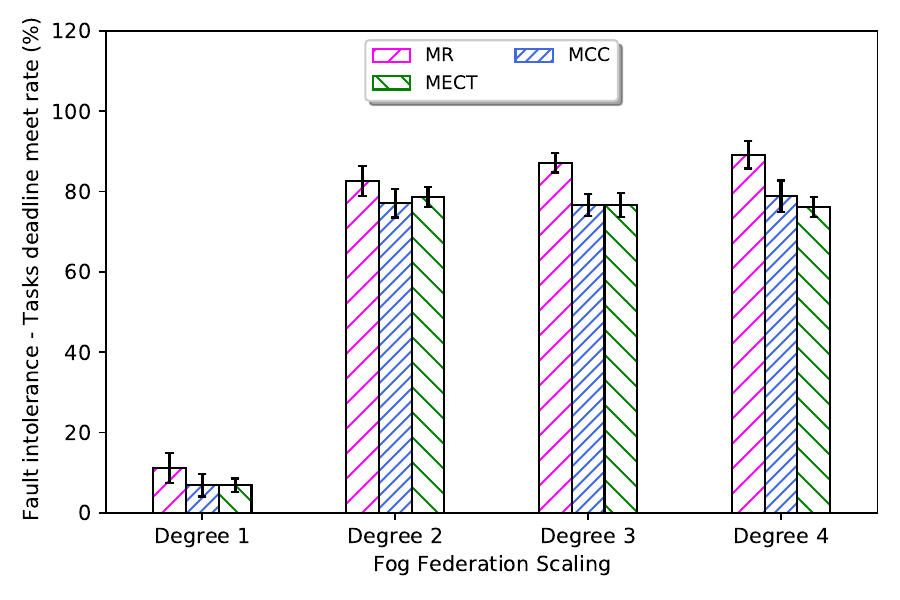}
    \caption{Impact of scaling the fog federation when different resource allocation methods are employed for monolithic applications on fault intolerance performance. The ``degree'' shows the number of neighbors the source fog system has in the federation.}
    \label{fig:fedScalingMon}
\end{figure}

To achieve our goal, we insert neighbouring fog systems ranging from one to four that correspond to degree one to four in the underlying fog federation graph. This means that, we develop a grid-based fog federation that can have minimum of one neighbour fog and maximum of four neighbour fog. To evaluate the system's performance, we capture the deadline meet rate of the requests under various oversubscription levels ranging from 100 to 400 workflow requests. This experiment utilizes the proposed ProPart partitioning and MR resource allocation methods. The result of this experiment is presented in Figure \ref{fig:fedScaling}. In addition, we performed a similar experiment for monolithic applications, where we fixed the number of receiving tasks to 1,000 requests and incremented the fog federation degree for various resource allocation methods. The result for the monolithic applications is presented in Figure~\ref{fig:fedScalingMon}.

\emph{Impact of Federation Scaling for Workflows on Fault Intolerance: }
Figure~\ref{fig:fedScaling} demonstrates the results for scaling the fog federation. We can see that, for any level of oversubscription, the largest federation (degree=4) excels. Importantly, considerable performance improvement is seen for the higher oversubscription level (300 and 400 workflows) that reflects the fault intolerance trend. On the other hand, for less overloaded cases, the performance improvement of scaling is marginal. Our analysis reveals that this is because ProPar attempts to put the entire workflow into one fog system, rather than partitioning and distributing them across the federation. For the higher levels of  oversubscription, however, the proposed method more often utilizes the fog federation to distribute workflow partitions. That is the reason we witness a more substantial improvement for fault intolerance performance in under higher oversubscription levels.

\emph{Impact of Federation Scaling for Monolith Applications on Fault Intolerance: }
In this experiment, we examine the resource allocation methods for monolithic applications while scaling the fog federation. As we can see in Figure \ref{fig:fedScalingMon}, the monolithic applications are positively impacted by the federation scaling. The result reflected a significant performance improvement when the federation scaled up from degree 1 to 2, regardless of the resource allocation method employed. Degree 1 describes a topology where there are only two fogs in the federation. That is why, none of the methods perform well. Even under such a tight resource availability, we can see that the MR method outperforms others. Nevertheless, for the higher degrees, MR performs approximately 15---18\% better than MECT and MCC. As mentioned earlier, the main reason behind superiority of MR is considering both the communication and computation latencies in its decision making. This particularly shows how the communication latencies can be decisive is such environments and that is why the and end-to-end latency has to be considered. 

\section{Conclusion and Future Works}\label{concl}
Transitioning to Industry 4.0, particularly at remote sites, entails the ability to handle modern micro-service workflows with strict latency constraints near data sources and in a serverless manner. The problem, however, is that under emergency situation, near-data fog computing systems quickly become oversubscribed and cannot handle the situation effectively. To overcome this challenge, we developed a platform for serverless industrial fog federation that is aware of the software architecture of the industrial applications and the characteristics (uncertainties) of the fog federation. The proposed platform operates within the gateways representing each fog system and makes fog federation serverless via hiding the topological complexities of the federation from the user's perspective. The platform's goal is achieved via two modules: the first one models each micro-service application as a DAG and tries to optimally partition it to subgraphs that can make use of the federation. Then, the second module, which is for resource allocation, maps the partitions across the fog federation via considering both computation and communication latencies. Evaluation results show the efficacy of the proposed platform, particularly under oversubscribed situations, where it brings about $\sim$15\%---18\% higher deadline meet rate in compare to other widely-used partitioning and resource allocation methods. Moreover, we noted that proposed solution for the federation is scalable and can dynamically handle different number of fogs in the federation. 

In the future, we plan to handle industrial use cases such as industrial monitoring operation (\eg drilling), that have to run uninterruptedly, across the federation. For that purpose, we are going to develop live service migration across the federation. In order to effectively manage industrial applications, it is imperative to support task priorities, such as urgent tasks with hard deadlines and best-effort tasks with soft deadlines (\cite{hujo2021toward,nouinou2023decision}). One complication is that some tasks of a certain industrial application can be considered as urgent, whereas, other tasks of the same application expose best-effort behavior. For the example of object detection application, the tasks used for fire detection (\cite{zhao2022fire}) are deemed urgent with hard deadlines, whereas, tasks for detecting employees' faces are best-effort and with lower priority. Demonstrating the feasibility of fog federation and partitioning micro-service workflows across it, one future avenue of this research can focus on the task characteristics and priorities across the fog federation. Another avenue for the future study will be performing partitioning in a more granular manner---within each micro-service. For instance, for ML-based micro-services, we can perform pre-processing on the source fog and neural network processing on another fog in the federation.

\section*{Acknowledgments}
 We would like to thank the anonymous reviewers of the paper. This research is supported by the National Science Foundation (NSF) under awards\# CNS-2047144 and IRES-Track1-2246390.

\printcredits

\balance
\bibliographystyle{cas-model2-names}

\bibliography{cas-refs}

\end{document}